\documentclass[]{article}

\usepackage{asp2004}
\usepackage{epsf}
\usepackage{psfig}
\usepackage{lscape}
\newcommand{\url}[1]{\texttt{\footnotesize{#1}}}
\begin{document}
\title{What do Multiple Planet Systems Teach us about Planet Formation?}   
\author{Eric B.\ Ford}   
\affil{University of California in Berkeley}    

\begin{abstract} 
For centuries, our knowledge of planetary systems and ideas about
planet formation were based on a single example, our solar
system. During the last thirteen years, the discovery of $\simeq170$
planetary systems has ushered in a new era for astronomy. I review the
surprising properties of extrasolar planetary systems and discuss how
they are reshaping theories of planet formation. I focus on how multiple planet
systems constrain the mechanisms proposed to explain the large eccentricities
typical of extrasolar planets.  I suggest that strong planet-planet scattering
is common and most planetary systems underwent a phase of large eccentricities.
I propose that a planetary system's final eccentricities may be strongly
influenced by how much mass remains in a planetesimal 
disk after the last strong planet-planet scattering event.
\end{abstract}

\section{Introduction}

For centuries, theories of planet formation had been designed to
explain our own Solar Systems, but the first few discoveries of
extrasolar planetary systems were wildly different than our own.
These discoveries led to the realization that planet formation theory
must be generalized to explain a much wider range of planetary
systems.  For example, traditional theories predicted that giant
planets would form at several AU and beyond, where temperatures are
cold enough for ices to initiate the growth of grains and
planetesimals (Lissauer 1993, 1995).  Now, we know of over 70 giant
planets inside 1 AU and 40 inside 0.1AU (http://www.obspm.fr/planets).
Theorists have proposed numerous possible mechanisms to explain the
existence of these planets.  Typically, they assume that the giant
planet formed beyond a few AU, but then migrated inwards through a
protoplanetary or planetesimal disk to their currently observed
locations (e.g., Goldreich \& Tremaine 1980; Lin et al.\ 1996;
Ward 1997; Murray et al.\ 1998; Cionco \& Brunini 2002) and stop before
being accreted on the star (e.g., Trilling et al.\ 1998; Ford \& Rasio
2006).  Similarly, it had long been assumed that planets formed in
circular orbits due to strong eccentricity damping in the
protoplanetary disk and remained on nearly circular orbits (i.e.,
eccentricity $\le$0.1; Lissauer 1993, 1995).  However, over half of
the extrasolar planets beyond 0.1AU have eccentricities $\ge$0.3, and
one is as large as $\simeq$0.95.  Theorists have suggested numerous
mechanisms to excite the orbital eccentricity of giant planets (e.g.,
Rasio \& Ford 1996;
Weidenschilling \& Marzari 1996; Lin \& Ida 1997; Holman et al.\ 1997;
Murray et al.\ 1998; Ford, Havlickova \& Rasio 2000; Kley 2000, 2004;
Chiang \& Murray 2002; Lee \& Peale 2002; Marzari \& Weidenschilling
2002; Ford, Rasio \& Yu 2003; Adams \& Laughlin 2003; Veras \&
Armitage 2004; Namouni 2005).  In recent years, improved observations
of a few multiple planet systems have allowed theorists to determine
their current orbital configuration and use that to place strong
constraints on the formation of a few planetary systems (Lee \& Peale
2002; Ford, Lystad \& Rasio 2005).

We review some of the mechanisms proposed to explain orbital migration in
disks in \S2 and eccentricity excitation in \S3.  In \S4, we review
the current knowledge of three particularly well-studied multiple
planet systems.  We conclude with a discussion of the implications of
these multiple planet systems for theories of orbital migration in
\S5.

\section{Orbital Migration}

\subsection{Interactions with Gaseous Disk}

Well before the discovery of extrasolar planets, analytic studies of a
planet in a gaseous protoplanetary disk indicated that torques could
lead to rapid orbital evolution (Goldreich \& Tremaine 1979, 1980).
Initially, it was not clear if the net torque would lead to inward or
outward migration, but subsequent investigations indicated that the
net torque typically leads to an inward migration for a single planet
in a quiet disk (Ward 1997).  Recently, numerous researchers have
conducted detailed hydrodynamic models to better understand the
details of the torques occurring at various locations in the disk.
While early work focused on torques exerted at Linblad resonances, it
is now clear that one must also consider torques occurring at
corotation resonances and accretion onto the planet, even once the
planet has cleared a gap in the disk (Artymowics \& Lubow 1996; Bate
et al.\ 2003; D'Angelo et al.\ 2003).  Unfortunately, these
complications demand that simulations include physics spanning a large
range of physical scales, and this remains a computational challenge.
While multiple groups have found qualitatively similar results, the
details remain a matter of active research (e.g., Bryden et al.\ 1999,
Kley 1999).  Further complicating matters, recent work has suggested
that turbulent fluctuations in the disk may be critical for
understanding migration (Rice \& Armitage 2003, Laughlin et al 2004).

Shortly after the discovery of giant planets in very short orbital
periods, it was realized that the planets likely formed at several AU,
but migrated to their current small orbital periods.  Torques from a
gaseous disk are widely believed to be responsible, as the torques
appear more than adequate to cause such large scale migrations.
Indeed, the main challenge to such theories is to explain why the
migration process is halted before the planet is accreted onto the
star.  Naively, one would expect the rate of migration to increase
with decreasing orbital period and the planets to accrete onto the
star.  Several halting mechanisms have been proposed (e.g., Trilling
1998), but it is not yet clear to what extent each of these mechanisms
is significant.  Many migration scenarios require some degree of fine
tuning (e.g., disk mass or lifetime) in order to halt the migration at
orbital periods of only 1.5-4d.

\subsection{Interactions with a Planetesimal Disk}

A disk of small solid bodies (e.g., protoplanets, planetesimals, pebbles)
can remain long after the gaseous protoplanetary nebula disperses
(Goldreich, Lithwick, Sari 2004).  If this disk is sufficiently
massive, then a giant planet could migrate through the disk by
scattering planetesimals (Murray et al.\ 1998; Cionco \& Brunini 2002;
Del Popolo \& Eks 2002).  Migration all the way to a few stellar radii
requires that the mass of planetesimal in the disk be large compared
to the observed disk masses of protoplanetary disks in Taurus and
Ophiuchus (Beckwith \& Sarget 1996).  Still, typical disk masses are
expected to result in a smaller amount of migration.  For a single
giant planet, the planetesimals that can be scattered at a given time
come from a relatively small range of semi-major axes near mean-motion
resonances, and the density of planetesimals must exceed a
significant threshold to power an extended period of migration.  When there
is more than one planet, the dynamics can become significantly more
complex and the feeding zones significantly enlarged.  For example, in
our own solar system, Saturn, Uranus, and Neptune are inefficient at
ejecting planetesimals, but efficiently scatter them inwards,
enabling Jupiter to eject them from the Solar System (Fernandez \& Ip
1984, Malhotra 1995).

\section{Eccentricity Excitation}

\subsection{Mutual Planetary Perturbations}

Mutual gravitation perturbations in multiple planet systems can 
lead to significant orbital evolution.  

\subsubsection{Secular Planetary Perturbations}

Secular perturbation theory approximates each planet as a ring of mass
smeared out over the planet's orbit.  In the secular approximation,
the semi-major axes remain constant, but the eccentricities,
inclinations, and orientations of the orbits evolve with time (Murray
\& Dermott 1999).  If the orbital planes are highly inclined
($\ge40^{\circ}$), then even a system with initially circular orbits
can undergo large eccentricity oscillations (the ``Kozai effect'';
Kozai 1962; Holman et al.\ 1997; Ford, Kozinsky \& Rasio 2000).  While
this effect is almost certainly important for some planets orbiting
stars that have a wide stellar binary companion, dissipation in the
protoplanetary disk makes it very unlikely for giant planets to form
with large relative inclinations (Lissauer 1993).  In the
low-inclinations and low-eccentricity regime, the eccentricity and
inclination oscillations decouple to lowest order, and angular
momentum is exchanged between the various planets on long timescales
(Murray \& Dermot 1999).  The low-inclination, high-eccentricity
regime can be studied by the octupole approximation (Ford, Kozinsky,
Rasio 2000; Lee \& Peale 2003) or by a numerical averaging procedure
(Michtechenko \& Malhotra 2004).  In both approximations, the
inclinations remain small, and conservation of angular momentum
requires that secular perturbations can only transfer angular momentum
from one orbit to another.  Therefore, secular planetary perturbations
can only excite significant eccentricities, if there is already at
least one eccentric planet in the system.

\subsubsection{Strong Planet-Planet Scattering}

If planet formation commonly results in planetary systems with
multiple planets, then it should be expected that the initial
configurations will not be dynamically stable for time spans orders of
magnitude longer than the timescale for planet formation (e.g.,
Levison, Lissuaer \& Duncan 1998).  When protoplanetary core form,
they do not know how much gas they will eventually accrete, so planets
will accrete too much mass to remain stable for the lifetime of their
star.  Additionally, giant planets must form while there is still
significant gas in the protoplanetary disk, so they are likely subject
to significant eccentricity damping which prevents eccentricity
growth.  Once the protoplanetary disk disperses, the eccentricity
damping is removed and mutual gravitational perturbations can start
exciting eccentricities that will eventually lead to close encounters.

In multiple planet systems which are dynamically unstable, close
encounters and strong planet-planet scattering can produce large
eccentricities (Rasio \& Ford 1996; Weidenschilling \& Marzari 1996).
For systems of two giant planets initially on nearly circular orbits,
dynamical instabilities are typically resolved by two planets
colliding and producing a more massive giant planet in another
low-eccentricity orbit or by one planet being ejected from the system,
leaving behind the other planet in an eccentric orbit.  For comparable
mass planets, this typically results in large eccentricities (Ford,
Havlickova \& Rasio 2001), but this same mechanism naturally produces
lower eccentricities when the planet mass ratio differs from unity
(Ford, Rasio \& Yu 2003).  While the distribution of eccentricities
depends on the planet mass ratio distribution, the two planet
scattering model predicts a maximum eccentricity of $\simeq0.8$,
independent of the mass ratio distribution.  This compares favorably
with the observed distribution of extrasolar planet eccentricities,
since only one of the $\simeq170$ known extrasolar planets has an
eccentricity greater than 0.8 (and the exceptional planet is in a
known binary).  The fraction of systems which result in ejections and
eccentric planets depends on the orbital distance and effective
radius for collisions (Ford, Havlickova \& Rasio 2001), as well as
the ratio of planet masses (Ford, Rasio \& Yu 2003).  While ejections
dominate for giant planets at several AU, collisions are more frequent
for comparable planets inside $\sim1$AU.  Therefore, strong
planet-planet scattering can easily produce the large eccentricities
of giant planets at large separations, but by itself would predict
that low-eccentricity orbits would be more frequent at small
separations.

Simulations of planet-planet scattering often begin with closely
spaced giant planets (e.g., Rasio \& Ford 1996; Ford, Havlickova \&
Rasio 2001).  This is necessary for dynamical instabilities to occur
in systems with only two planets initially on circular orbits.  While
such systems facilitate the systematic study of the relevant physics,
real planetary systems likely have more than two massive bodies.  In
planetary systems with multiple planets, dynamical stabilities are
common even for systems with large initial separations (Chambers,
Wetherill \& Boss 1996; Marzari \& Weidenschilling 2002).
Additionally, such systems can persist uneventfully for
$\sim10^{6-8}$yr, before chaos leads to close encounters and strong
planet-planet scattering.

\subsubsection{Dynamical Relaxation}

If protoplanetary disks form many planets nearly simultaneously,
then planet-planet scattering may lead to a phase of dynamical
relaxation.  Several researchers have numerically investigated the
dynamics of planetary systems with ~10-100 planets (Lin \& Ida 1997;
Papaloizou \& Terquem 2001, 2002; Adams \& Laughlin 2003; Barnes \&
Quinn 2004).  Initially, such systems are highly chaotic and close
encounters are common.  The close encounters lead to planets colliding
(creating more massive planet) and/or planets being ejected from the
system, depending on the orbital periods and planet radii.  Either
process results in the number of planets in the system being reduced
and the typical separations between planets increasing.  The system
gradually evolves from a rapidly unstable state to states which will
endure longer before the next collision or ejection.  Such systems
typically evolve to a state with 1-3 eccentric giant planets which
will persist for the lifetime of the star.  In systems with at least
two remaining planets, the typical ratio of semi-major axes of the
innermost planets is typically large, but show considerable variation
across different systems, $\left<a_2/a_1\right> = 25\pm24$ and
$11\pm7.8$ for two different mass distributions (Table.~4 of Adams \&
Laughlin 2003).  These distribution of final eccentricities in such
systems displays a breadth comparable to the observed distribution of
eccentricities of extrasolar planets, but underproduce planets with
small eccentricities.  Although dynamical relation does not predict a
strict upper limit for the eccentricities generated (as does
planet-planet scattering with two planets initially on circular
orbits), extreme eccentricities are unlikely ($p(e>0.8)\le0.1$; see
Fig.~7 of Adams \& Laughlin 2003), since the final eccentricities are
the result of a succession of ejections and/or collisions.

Since the initial evolution is strongly chaotic, the results of such
simulations are relatively insensitive to the exact choice of initial
conditions, but bounded by conservations of energy and angular
momentum.  This partially explains the similar results of several
groups using different initial conditions.  However, nearly all such
simulations have considered purely gravitational forces.  In fact,
planetary systems may evolve via dynamical relaxation while the disk
still has a significant amount of mass in gas or planetesimals.
Either a gas or planetesimal disk is likely to provide a significant
amount of dissipation which could significantly alter the evolution of
the system.  While some work has investigated the effects of
dissipative gaseous disk which drives convergent migration between two
planets and lead to close encounters (Adams \& Laughlin 2003; Moorhead
\& Adams 2005), much more work remains to be done to explore the wide
range of parameter space which exists for systems with multiple
planets and a dissipative disk.

\section{Three Multiple-Planet Systems}

First, we review recent research on the history of three well-studied
multiple planet systems orbiting three solar type stars: the Sun, GJ
876, and Upsilon Andromedae ($\upsilon$ And).  Several other multiple planet systems
have been discovered by radial velocity searches, but either the planets
interact too weakly to provide dynamical constraints on planet formation
or the published
observations are not yet sufficient to precisely constrain their
dynamics.  Even though high precision measurements are also available
for the planets orbiting pulsar PSR 1257+12, we do not include this
system, since it's formation may have been very different than planet
formation around solar type stars.

\subsection{The Solar System}

Despite centuries of study and {\em it situ} measurements by space
probes, the formation of giant planets in our solar system remains a
matter of significant debate.  In particular, it is not certain
whether giant planets form via the gradual accretion of a rocky core
or via direct gravitational collapse.
According to the gravitational instability model, giant planets are
formed by gravitational instabilities in the protoplanetary disk, much
like binary stars (Boss 1995, 1996).  These simulations are very
computationally challenging, so they are not able to include all the
relevant physics.  Whether or not giant planets form depends on the
simplifying assumptions used for the simulation.  While some numerical
simulations form massive giant planets in a few orbital times, these
typically start from disks that are violently unstable.  Further,
these typical integrations are run for such a short period of time
that they can not start from plausible initial conditions.  Recent
simulations have considered disks that start from a stable state and
gradually approach instability via cooling (Pickett et al.\ 2003;
Mejia et al.\ 2005).  These simulations form rings and can temporarily
fragment, if the cooling time is sufficiently rapid, but they have not
resulted in forming stable giant planets.  In principle, the main
advantage of the gravitational instability model is that it might be
able to form giant planets rapidly, even at large orbital separations.
Another potential advantage is that the giant planets would typically
be formed in eccentric orbits.  Thus, the significant eccentricities
of extrasolar planets could be explained without invoking any
additional mechanisms for eccentricity excitation.

According to the competing model of core accretion, collisions between
rocky planetesimals result in the gradual growth of a rocky core
(Lissauer 1993).  Once the core becomes sufficiently massive, it
accretes a large quantity of gas from the protoplanetary disk (Pollack
et al.\ 1996).  Several details of this model remain active areas of
research (e.g., ``Why do collisions between planetesimals result in
accretion rather than shattering?'' and ``How do small planetesimals
avoid rapid orbital decay in the protoplanetary disk?'').  Still,
there is little doubt that core accretion must explain the formation
of the terrestrial planets, asteroids, and other small bodies in the
solar system.  However, there is active debate whether core accretion
could have formed the cores of Uranus and Neptune before the gas disk
dissipated.  This has led some researchers to propose that Uranus and
Neptune, and perhaps all four giant planets, may have formed via
gravitational instability.  Other researchers have proposed
refinements to the core accretion model that could allow for the more
rapid formation of Uranus and Neptune.  Here we summarize two recent
attempts to explain the formation of Uranus and Neptune within the
core accretion framework.

Two similar scenarios for forming Uranus and Neptune via core
accretion both suggest that they initially formed at much smaller
orbital distances, where the timescales relevant for planet formation
are shorter.  In one version, Thommes, Duncan \& Levison (1999)
proposed that Uranus and Neptune formed much closer to the Sun than
their current orbital separations, perhaps even between Jupiter and
Saturn.  As the disk began to dissipate, planet-planet scattering
excited large eccentricities and caused their orbits to extend well
beyond Saturn.  Then dynamical friction in the protoplanetary disk
would have circularized their orbits at orbital separations comparable
to those we see today.  In a slightly refined version, Uranus and
Neptune again would have initially formed closer to the Sun than their
current orbital separations (but still beyond Saturn).  This closely
packed system could survive for an extended period of time if the
eccentricities of all four giant planets were significantly smaller
than they are today.  Planetesimal scattering would have caused
Jupiter to migrate slightly inwards, while Saturn, Uranus, and Neptune
would have migrated outwards. The eccentricities would have remained
small until Saturn crossed the 2:1 mean motion resonance with Jupiter.
This divergent resonant crossing would not result in resonance
capture, but would excite significant eccentricities that would
propagate throughout the system.  Uranus and Neptune would be
scattered outwards, but could have circularized near their current
orbits due to dynamical friction with a planetesimal disk (Fig.\ 1;
Tsiganis et al.\ 2005).  This scenario is particularly appealing,
since n-body simulations show that it can also reproduce several other
observed properties of the solar system (Morbidelli et al.\ 2005;
Gomes et al.\ 2005; Strom et al.\ 2005).

Another possibility is that a collisional cascade maintained a
significant fraction of the disk mass in small rocky bodies, even after
protoplanets have formed (Goldreich, Lithwick \& Sari 2004).  In this
scenario, several Uranus and Neptune-mass protoplanets could have
formed near the current location of Uranus and Neptune, since
dynamical friction damped the random velocities and gravitational focusing allowed them to
accrete more rapidly than conventionally assumed in the core accretion
model.  Eventually, the mass in the small bodies must have decreased
to the point where dynamical friction was no longer sufficient to
prevent the protoplanets from exciting each other's eccentricities.
Then the protoplanets could have close encounters and scatter each
other.  In the solar system, several massive proto-planets would have
been scattered from near Neptune inward to Uranus, then on to Saturn
and Jupiter, before being ejected from the Solar System.  Once Uranus
and Neptune were the only remaining massive bodies remaining, both
planets would be expected to have large eccentricities from scattering
nearly comparable mass protoplanets inwards (Chambers 2001).
Therefore, some mechanism for eccentricity damping would be necessary
to explain their current low eccentricity orbits.  The circularization
could be caused by dynamical friction and planetessimal scattering in
what remains of the planetessimal disk.

\subsection{GJ 876}

Three planets have been discovered around the M4 dwarf, GJ 876 (Marcy
et al.\ 2001; Rivera et al.\ 2005).  The most recently discovered planet
(d) has a minimum mass of $\simeq6 M_\oplus$ planet and orbits at
$0.02$AU, but is not essential for our subsequent discussion of the
orbital evolution of the outer two planets.  The two more massive
planets (b \& c) have minimum masses of 1.9 and 0.6 $M_{\mathrm{Jup}}$
and orbit at 0.21 and 0.13AU, respectively.  The middle planet has an
eccentricity $\simeq0.2$, but the outer planet's eccentricity is much
smaller ($\le0.03$).  These planets are particularly interesting,
since they are near a 2:1 mean motion resonance, and mutual
planetary perturbations have already been observed (Laughlin et al.\
2005).

Since mean motion resonances occupy only a small fraction of the
available phase space, one might naively assume that it is unlikely
for two planets to form in a mean motion resonance.  However, if
significant planetary migration and multiple planet systems are both
common, then planets could form away from mean motion resonances and
differential migration could cause the planets approach a mean motion
resonance.  If the migration is both smooth and convergent, then as
planets approach mean-motion resonances, they can be efficiently
captured into a low-order mean-motion resonance.  Thus, the pair of
planets in GJ 876 suggests that significant migration is likely to
have occurred in that planetary system.
If the migration were to continue after resonant capture, then both planets
would migrate together, leading to significant eccentricity evolution
(Peale 1986).  Indeed, hydrodynamic simulations of two planets
embedded in a gaseous disk confirm this behavior (Bryden et al.\ 2000;
Kley 2000; Snellgrove, Papaloizou, Masset \& Nelson 2001; Papaloizou
2003; Kley et al.\ 2005) Therefore, eccentricity excitation via
resonance capture is a natural explanation for the observed
eccentricities for those extrasolar planetary systems which
participate in low-order mean-motion resonances.  This possibility has
been studied intensively in the context of GJ 876 (Lee \& Peale 2002;
Snellgrove, Papaloizou \& Nelson 2001; Kley et al.\ 2005), as well as
extrasolar planetary systems more generally (Lee 2004; Nelson \&
Papaloizou 2002).

Lee \& Peale (2002) studied the evolution of GJ 876b \& c, assuming
initially well-separated circular orbits and a smooth convergent
migration leading to capture in the 2:1 mean motion resonance.  This
naturally leads to eccentricity excitation and can easily generate the
observed eccentricity of planet c and a small eccentricity for planet
b.  In fact, the eccentricity excitation due to resonance capture is
so efficient that this places significant constraints on the migration
history.  In one scenario, the migration would have led to capture in
the 2:1 mean motion resonance, but the migration must have halted very shortly
afterwards.  Lee \& Peale (2002) estimate that the semi-major axis of
the outer planet could only decrease by 7\% after resonance capture,
requiring the protoplanetary nebula to dissipate at nearly the same
time as the capture into resonance.  Since this scenario would require an
unlikely fine-tuning of parameters, they develop an alternative model
which includes eccentricity damping due to interactions with the disk
of the form $\dot{e}/e = -K \dot{a}/a$, where $e$ is the eccentricity,
$a$ is the semimajor axis, the dots represent time derivatives, and
$K$ is a numerical constant.  Significant eccentricity damping could
slow the excitation of eccentricities and allow the planets to migrate
by more than 7\% after the resonance capture, somewhat reducing the
level of fine-tuning needed.  If $K\sim100$, then the eccentricities
start to grow after resonance capture, but saturate at near the
currently observed eccentricities, eliminating the need for the
migration to halt shortly after resonance capture (see Fig.\ 2, left).
More detailed hydrodynamic simulations confirm this finding
(Papaloizou 2003; Kley et al.\ 2004; Kley et al.\ 2005).  Kley et al.\
2005 used the revised orbital fits from Laughlin et al.\ (2004)
and found that $K\simeq40-170$ was needed for the
eccentricity excitation to saturate near the current values, depending on the
inclination of the system (but assuming coplanar orbits with an inclination relative to the plane of the
sky greater $35^\circ$, as suggested by radial velocity constraints).

\subsection{$\upsilon$ Andromedae}

The system of three giant planets orbiting $\upsilon$ And (Butler
et al.\ 1999) also offers clues to the history of orbital migration.
Like GJ876, one planet (b) has a short orbital period (4.6d) and is
not essential for understanding the dynamics of the outer two planets.
The outer two planets (c \& d) have orbital periods of 241d and 1301d
and eccentricities of 0.26 and 0.28, respectively (Ford, Lystad \&
Rasio 2005).  Soon after their discovery, it was realized that mutual
planetary perturbations could cause significant secular evolution of
the eccentricities and longitudes of periastron for the outer two
planets (Stepinsky, Malhotra \& Black 2000; Chiang, Tabachnik \&
Tremaine 2001; Lissauer \& Rivera 2001).

Two models were proposed to explain the eccentricities and longitudes
of pericenter for planets c \& d.  Chiang \& Murray (2002) proposed
that a protoplanetary disk beyond planet d could {\em adiabatically}
torque planet d.  If the longitudes of periastron were initially
circulating, then this torque would drive the system towards solutions
where the longitudes of periastron librate about an aligned
configuration.  Once the system was in the librating regime, the
torque would damp the libration amplitude.  Thus, this model would
predict that the the pericenters of the outer two planets would
currently be librating with small amplitude about an aligned
configuration and that the secular evolution would cause only small
variations in the eccentricities.
Malhotra (2002) proposed an alternative model in which the outer
planet was perturbed {\em impulsively}, as would be expected if it had
a close encounter with another (undetected) planet.  In this scenario,
the two planets could be either librating or circulating, depending on
the relative phases at the time of the impulsive perturbation.  If the
system were librating, then this model would generally predict that
the libration amplitude would be large and that there would be
significant eccentricity oscillations.

The best-fit orbital solution to the early observations suggested that
the pericenter directions of the outer two planets were very nearly
aligned ($\le 10^\circ$; Butler et al.\ 1999), favoring the model for
adiabatic perturbations from a disk (Chiang \& Murray 2002).  However,
subsequent observations show that the pericenters are less well
aligned than previously thought ($\Delta \omega =
37.6^{\circ}\pm4.8^{\circ}$; Ford, Lystad \& Rasio 2005), favoring an
impulsive perturbation due to planet scattering.

The planetary system around $\upsilon$ And has an even more remarkable
property.  This system lies very close to the boundary between librating and circulating solutions.  As a result, 
%
%
the eccentricity of the middle planet undergoes very
large oscillations with $e$ ranging from from 0.34 to very nearly zero
(see Fig.\ 2, right, after $10^4$ years).  Stepinsky, Malhotra \&
Black (2000) recognized that this was possible for {\em some} orbital
solutions consistent with the radial velocity observations.  Ford,
Lystad \& Rasio (2005) used a rigorous Bayesian statistical analysis
to demonstrate that the eccentricity of the middle planet 
periodically returns to nearly zero for {\em all allowed} orbital
solutions (see Fig.\ 2 of Ford, Lystad \& Rasio 2005).  This provides
a strong constraint on the timescale for eccentricity excitation in
$\upsilon$ And ($\simeq100$yr).  For a planet-disk interaction to
excite an eccentricity of $\simeq0.3$ would require a very massive
disk ($\ge 40 M_{\mathrm{Jup}}$) exerting a very strong torque only to abruptly
stop after $\simeq100$yr.  Thus, this peculiar orbital configuration
would be extremely unlikely, unless both planets were initially on
circular orbits and the outer planet were perturbed impulsively by
strong planet-planet scattering (Malhotra 2002).


\section{Implications for Planet Formation}

\subsection{Orbital Migration}

Regardless of how the giant planets formed, the large number of Kuiper
belt objects in mean-motion resonances with Neptune provides strong
evidence for significant outward migration of Neptune via planetesimal
scattering (Hahn \& Malhotra 1999).  Numerical simulations have shown
that the necessary migration is naturally explained via planetesimal
scattering for reasonable disk masses.  Only Jupiter is efficient at
ejecting planetesimals from the Solar System, but together Neptune,
Uranus, and Saturn can scatter planetesimals from near Neptune's orbit
inwards to Jupiter.  Therefore, Jupiter migrated inwards (slightly due
to its large mass), while Saturn, Uranus, and Neptune migrated
outwards (Fernandez \& Ip 1984; Malhotra 1995).

Initially, theoretical difficulties for forming giant planets at
orbital separations of $\simeq0.05$AU helped rekindle models of planet
migration.  The detection of pairs of planets in 2:1 mean motion
resonances (e.g., GJ 876 b \& c) suggests that smooth convergent
migration likely occurred in these systems.  Additionally, the fact that
migration models can simultaneously match the observed eccentricities
for both planets b \& c provides further evidence for migration in
this system.

\subsection{Eccentricity Excitation via Orbital Migration} 

It is natural to ask if the large torques presumed responsible for
orbital migration could also be responsible for exciting orbital
eccentricities.  

\subsubsection{Migration in Planetesimal Disk}

Analytical arguments suggest that the
planetesimals typically provide a source of dynamical friction
(Goldreich et al.\ 2004).  Simulations of a single-planet scattering
planetesimals in the Opik approximation also show that eccentricities
are usually damped (Murray et al.\ 1998), although eccentricity
excitation may be possible for sufficiently massive planets ($\ge$10
$M_{\mathrm{Jup}}$).  In our own solar system, it is also believed that
scattering of planetesimals may have damped the eccentricities of the
outer planets after violent events.  Finally, direct simulations of our
solar system also demonstrate that planetesimal scattering typically
damps eccentricities (Hahn \& Malhotra 1999; Thommes, Duncan \&
Levison 1999, 2002; Tsiganis et al.\ 2005).  

\subsubsection{Migration in Gaseous Disk}

While the dissipative nature of a gaseous disk naturally leads to
eccentricity damping (Artymowicz 1993), a few researchers have
suggested that excitation may also be possible.  Artymowicz (1992)
found that a sufficiently massive giant planet ($\ge$10 $M_{\mathrm{
Jup}}$) can open a wide gap, leading to torques which excite
eccentricities.  More recently, Goldreich \& Sari (2003) have
suggested that a gas disk could excite eccentricities even for less
massive planets via a finite amplitude instability.  This claim is
controversial, as 3-d numerical simulations have not been able to
reproduce this behavior (e.g., Papalouizou et al.\ 2001; Ogilvie \&
Lubow 2003).  Given the large dynamic ranges involved and the 
complexity of the simulations, one might question the accuracy of
current simulations.  For example, three dimensional simulations have
suggested that the gaps induced by giant planets might not be as well
cleared as assumed in many two dimensional disk models (Bate et al.\ 2003;
D'Angelo et al.\ 2003).  We believe that further theoretical and
numerical work is needed to better understand planet-disk
interactions.  In the mean time, we look to the observations for
guidance on the question of eccentricity damping or excitation.

\subsubsection{Empirical Evidence}

In the GJ876 system, the observed eccentricities are not consistent
with eccentricity excitation via interactions with the disk.  The
current observed eccentricities could be readily explained if
interactions with a gas disk led to strong eccentricity damping $K =
\dot{e} a / e \dot{a} \gg1$ (Lee \& Peale 2002; Kley et al.\ 2005).
This is in sharp contrast to current hydrodynamic simulations of
migration that suggest $K\simeq1$ and theories that predict $K<0$
(e.g., Goldreich \& Sari 2003; Ogilvie \& Lubow 2003).  While other
planetary systems are not yet as well constrained or studied as GJ
876, the moderate eccentricities of other extrasolar planetary systems
near the 2:1 mean motion resonance suggest that GJ 876 is not unique.

The $\upsilon$ And system also provides a constraint on
eccentricity excitation during migration.  If the outer two planets
migrated to their current locations (0.8 and 2.5AU), then they must
have been in nearly circular orbits at the time of the impulsive
perturbation in order for the middle planet's eccentricity to
periodically return to nearly zero.  While this does not demonstrate a
need for rapid eccentricity damping as in GJ 876, this is inconsistent	
with models which predict significant eccentricity excitation.  Since
dynamical analyses severely limit the possibility of eccentricity
excitation in both the GJ 876 and $\upsilon$ And systems, we
conclude that orbital migration does not typically excite eccentricities, at
least for a planet-star mass ratio less than $\sim0.003-0.006$ (those
of the most massive planet in $\upsilon$ And and GJ 876).

\subsection{Origin of Eccentricities}

Empirical constraints that suggest that interactions with a gaseous
disk do not typically excite eccentricities (\S5.1).  For GJ 876 (and other
planetary systems near mean motion resonances)
continued migration after resonance capture 
could excite the eccentricities of the outer two planets.  However, this
mechanism is insufficient for explaining the eccentricities of extrasolar
planets in general, since the majority of observed multiple planet systems are
not near a low-order mean-motion resonance.  The dramatic eccentricity
oscillations of $\upsilon$ And c provide an upper limit on the
timescale for eccentricity excitation in $\upsilon$ And ($\simeq
100$yr) and strong evidence for planet-planet scattering in this
system (Ford, Lystad \& Rasio 2005).
%
%
Planet-planet scattering in either few-planet systems (Ford, Rasio \& Yu 2003) or
many-planet systems (Adams \& Laughlin 2003) could produce an
eccentricity distribution quite similar to that observed for
extrasolar planets.
%

A complete theory of planet formation must explain both the eccentric
orbits prevalent among extrasolar planets and the nearly circular
orbits in the Solar System.  Despite significant uncertainties about
giant planet formation, all three mechanisms for forming the Solar
System's giant planets (see \S4.1) agree that the giant planets in the
Solar System went through a phase of large eccentricities.  If Uranus
and Neptune formed closer to the Sun, then close encounters are
necessary to scatter them outwards to their current orbital distances.
During this phase, their eccentricities can exceed $\simeq0.5$
(Tsiganis et al.\ 2005).  Alternatively, if Uranus and Neptune were
able to form near their current locations due to eccentricity damping
from a disk of small bodies, then several other ice giants should have
formed contemporaneously in the region between Uranus and Neptune.
The scattering necessary to to remove these extra ice giants would
have excited sizable eccentricities in Uranus and Neptune (Goldreich,
Lithwick \& Sari 2004).  Finally, the gravitational instability model
predicts that most giant planets form with significant eccentricities.
Therefore, it seems most likely that even the giant planets in our
Solar System were once eccentric.


Perhaps the question, ``What mechanism excites the eccentricity of
extrasolar planets?'' should be replaced with ``What mechanism damps
the eccentricities of giant planets?''  Unless giant planets form via
gravitational instability, interactions with a gas disk are not an
option, since the eccentricities would have been excited after the gas
was cleared.  Both dynamical friction within a planetesimal disk and
planetesimal scattering could damp eccentricities in both the Solar
System and other planetary systems.  Dynamical friction alone would
not clear the small bodies, so either accretion or ejection
would be required to satisfy observational constraints (Goldreich,
Lithwick \& Sari 2004).  Planetesimal scattering provides a natural
mechanism to simultaneously damp eccentricities and remove 
small bodies from planetary systems.  

Perhaps, {\em the key parameter that determines whether a planetary
system will have eccentric or nearly circular orbits is the amount of
mass in planetesimals at the time of the last strong planet-planet
scattering event.}
The chaotic evolution of multiple planet systems naturally provides a
large dispersion in the time until dynamical instability results in
close encounters (Chambers, Wetherill \& Boss 1996; Ford, Havlickova 
\& Rasio 2001; Marzari \& Weidenschilling 2002).  
Unfortunately, this could significantly complicate the
interpretation of the observed eccentricity distribution for
extrasolar planets.  
On a positive note, this model might naturally explain both the
eccentric orbits of extrasolar planets and the circular orbits in the
Solar System.  Future numerical investigations will be necessary
to test this model further.

\acknowledgements E.B.F. thanks E.I. Chiang, G. Laughlin, M.H. Lee, H. Levison,
G.W. Marcy, A. Morbidelli, J.C.B. Papaloizou, S. Peale, F.A. Rasio, and J. Wright 
for useful discussions.  
E.B.F. acknowledges the support of the Miller Institute for Basic Research.


\newpage

%
\noindent See \url{http://www.nature.com/nature/journal/v435/n7041/fig\_tab/nature03539\_F1.html}.  \\
\vspace{0.01in} \\
Figure 1:  
Orbital evolution of a hypothetical planetary system similar to the
Solar System.  The lines show the semimajor axis (middle lines),
periastron distance (q; lower lines), and apastron distance (Q; upper
lines) for each planet.  This n-body simulation started with the giant
planets closer together than the Solar System giant planets are today.
The planets migrated due to scattering planetesimals from a 35
$M_\oplus$ disk extending out to 30 AU.  The vertical dotted line
marks the epoch where Jupiter and Saturn crossed their 1:2 mean motion
resonance.  After this point, large eccentricities were excited and
the planets underwent close encounters and strong planet-planet
scattering.  For example, the orbits of planets U and N cross.
Continued planetesimal scattering damps the eccentricities to near the
present values for the solar system giant planets.  The values at the
right indicate the maximum eccentricities of each planet over the last
2Myr. 
Reprinted by permission from Macmillan Publishers Ltd: {\em Nature}
(Tsiganis et al.\ 2005), copyright 2005. \\
%
%
%
\vspace{0.5in} \\
%
%
\noindent See \url{http://www.journals.uchicago.edu/ApJ/journal/issues/ApJ/v567n1/54571/54571.figures.html}.  \\
\vspace{0.01in} \\
Figure 2:  
Model for the eccentricity evolution of the outer two planets in
GJ 876 due to smooth convergent migration.  The solid curves show how
the eccentricities are excited following capture into the 2:1 mean
motion resonance for different assumptions about the rate of
eccentricity damping.  The horizontal dashed lines show the
approximate observed eccentricities for the planets.  Unless there was
strong eccentricity damping, continued migration after capture into
the 2:1 mean motion resonance would rapidly cause the eccentricities
to exceed their observed values.  In this model, an outer disk is
assumed to torque only the outer planet.  More sophisticated models
give similar results (e.g., Kley et al.\ 2005).  Note that the inner
planet is referred to as 1 in the figure and c in the text, and the
outer planet is referred to as 2 in the figure and b in the text.
Reproduced by the kind permission of the AAS (Lee \& Peale 2002). 
\\
%
\vspace{0.5in} \\
%
\noindent See \url{http://www.nature.com/nature/journal/v434/n7035/fig\_tab/nature03427\_F4.html}.  \\
\vspace{0.01in} \\
Figure 3:
Dynamical evolution of a hypothetical planetary system similar
to $\upsilon$ And.  The top panel shows the semimajor axes (middle
lines) and peri and apastron distance (lower and upper lines) for
planets similar to the middle (C, dashed line) and outer (D, dotted
line) planets around $\upsilon$ And, as well as a hypothetical fourth
planet (E, solid line).  The innermost planet, B, is not shown, as it
plays a negligible role.  The lower panel shows the eccentricity
evolution for the same numerical integration.  After a brief period of
dynamical instability, planet E is ejected, leaving the other two in a
configuration that is very similar to that presently observed for
$\upsilon$ And c and d.  
Note that the timescale to completely eject the outer planet from the system (after $\simeq$9,000 years in
this particular simulation) is much longer than the timescale of the initial strong
scattering ($\simeq$100 years).  After this initial brief phase of strong interaction, the perturbations on the
outer planet are too weak to affect significantly the coupled secular evolution of $\upsilon$ And C
and D. Thus, the ``initial'' eccentricity of $\upsilon$ And C for the secular evolution is determined by its value at the end of the strong interaction phase, rather than that at the time of the final ejection.
Reprinted by permission from Macmillan Publishers Ltd: {\em Nature} (Ford, Lystad \& Rasio 2005), copyright 2005.
%


\begin{thebibliography}{}


\bibitem[]{} Adams, F.C., \& Laughlin, G. 2003 Icarus 163, 290.
\bibitem[]{} Artymowicz 1992 PASP 104, 769.
\bibitem[]{} Artymowicz 1993 ApJ 419, 116.
\bibitem[]{} Artymowicz, P. Lubow, S.H. 1996 ApJ 476, L77.
\bibitem[]{} Barnes, R. Qunn, T. 2004 ApJ 611, 494.
\bibitem[]{} Bate, M.R., Lubow, S.H., Ogilvie, G.I., Miller, K.A. 2003 MNRAS 341, 213.
\bibitem[]{} Beckwith, S.V.W. \& Sargent, A.I. 1996 Nature 383, 189.
\bibitem[]{} Boss, A.P. 1995 Science, 267, 360.
\bibitem[]{} Boss, A.P. 1996 L\&PS, 27, 139.
\bibitem[]{} Bryden, G., Chen, X., Lin, D.C.N., et al.\ 1999 ApJ 514, 334.
\bibitem[]{} Bryden, G., R\'ozyczka, M., Lin, D.N.C. \& Bodenheimer, P. 2000 ApJ, 540, 1091
\bibitem[]{} Butler, R.P., et al.\ 1999 ApJ, 526, 916.
\bibitem[]{} Chambers, J.E. 2001 Icarus 152, 205.
\bibitem[]{} Chambers, J.E., Wetherill, G.W. \& Boss, A.P. 1996 Icarus 119, 261.
\bibitem[]{} Chiang, E.I. 2003 ApJ 584, 465.

\bibitem[]{} Chiang, E.I., Fischer, D., Thommes, E. 2002 ApJ 564, L105.
\bibitem[]{} Chiang, E.I. \& Murray, N. 2002 ApJ 576, 473.
\bibitem[]{} Chiang, E.I., Tabachnik, S, Tremaine, S. 2001 AJ 122, 1607.
\bibitem[]{} Cionco, R.G. \& Brunini, A. 2002 MNRAS, 334, 77.
\bibitem[]{} D'Angelo, G., Kley, W, Henning, T. 2003 ApJ 586, 540.
\bibitem[]{} Del Popolo, A., Eks, I.,K.Y. 2002 MNRAS 332, 485.
\bibitem[]{} Fernandez, J.A., Ip, W.-J. 1984 Icarus 58, 109.
\bibitem[]{} Ford, E.B., Havlickova, M. \& Rasio, F.A. 2001 Icarus 150, 303.
\bibitem[]{} Ford, E.B., Kozinsky, B., Rasio, F.A. 2000 ApJ 535, 385.
\bibitem[]{} Ford, E.B., Lystad, V., Rasio, F.A. 2005 Nature 434, 873.
\bibitem[]{} Ford, E.B. \& Rasio, F.A. 2006, submitted to ApJL.
\bibitem[]{} Ford, E.B., Rasio, F.A., Yu, K. 2003 Scientific Frontiers in Research on Extrasolar Planets, eds. D. Deming \& S. Seager (ASP Conference Series, 294), 181.
\bibitem[]{} Goldreich, P., Lithwick, Y., Sari, R. 2004 ApJ 614, 497.
\bibitem[]{} Goldreich, P., Sari, R. 2003 ApJ 585, 1024.
\bibitem[]{} Goldreich, P., Tremaine, S. 1979 ApJ 233, 857.
\bibitem[]{} Goldreich, P., Tremaine, S. 1980 ApJ 241, 425.
\bibitem[]{} Gomes, R., Levison, H.F., Tsiganis, K., Morbidelli, A. 2005 Nature 435, 466.
\bibitem[]{} Hahn, J.M., Malhotra, R. 1999 AJ 117, 3041.
\bibitem[]{} Holman, M., Touma, J., Tremaine, S. 1997 Nature 386, 254.
\bibitem[]{} Kley, W. 1999 MNRAS 303, 696.
\bibitem[]{} Kley, W. 2000 MNRAS 313, L47.
\bibitem[]{} Kley, W., Lee, M.H., Murray, N., Peale, S.J. 2005 A\&A 437, 727.
\bibitem[]{} Kley, W., Peitz, J. \& Bryden, G. 2004 A\&A, 414, 735.
\bibitem[]{} Kozai, Y. 1962 AJ 67, 591.
\bibitem[]{} Laughlin, G., Butler, R.P., Fischer, D.A., Marcy, G.W., Vogt, S.S., Wolf, A.S. 2005 ApJ 622,1182. 
\bibitem[]{} Laughlin, G., Steinacker, A., Adams, F.C. 2004 ApJ 608, 489.
\bibitem[]{} Lee, M.H. 2004 ApJ 611, 517.
\bibitem[]{} Lee, M.H. \& Peale, S.J. 2002 ApJ 567, 596.
\bibitem[]{} Lee, M.H. Peale, S.J. 2003 ApJ 592, 1201.
\bibitem[]{} Levison, H.F., Lissuaer, J.J., Duncan, M.J. 1998 AJ 116, 1998.
\bibitem[]{} Lin, D.N.C., Bodenheimer, P., Richardson, D.C. 1996 Nature 380, 606.
\bibitem[]{} Lin, D.N.C. \& Ida, S. 1997 ApJ, 447, 781
\bibitem[]{} Lissauer, J.J. 1993 ARAA, 31, 129
\bibitem[]{} Lissauer, J.J. 1995 Icarus 114, 217.
\bibitem[]{} Lissauer, J.J. \& Rivera, E.J. 2001 ApJ 554, 1141. 
\bibitem[]{} Malhotra, R. 1995 AJ 110, 420.
\bibitem[]{} Malhotra, R. 2002 ApJ, 575, 33.
\bibitem[]{} Marcy, G., et al.\ 2001 ApJ, 555, 418.
\bibitem[]{} Marzari, F. \& Weidenschilling, S.J. 2002 Icarus 156, 570.
\bibitem[]{} Mejia, A.C., Durisen, R.J., Pickett, M.K., Cai, K. 2005 ApJ 619, 1098.
\bibitem[]{} Michtechenko, R.A., Malhotra, R. 2004 Icarus 168, 237.
\bibitem[]{} Morbidelli, A., Levison, H.F., Tsiganis, K., Gomes, R. 2005 Nature 435, 462.
\bibitem[]{} Murray, C.D. \& Dermott, S.F. 1999 Solar System Dynamics (New York: Cambridge University Press)
\bibitem[]{} Murray, N. Hansen, B., Holman, M., Tremaine, S. 1998 Science 279, 69.
\bibitem[]{} Nagasawa, M., Lin, D.N.C., Ida, S. 2003 ApJ 586, 1374.
\bibitem[]{} Namouni, F. 2005 AJ 130, 280. 
\bibitem[]{} Nelson, R.P. \& Papaloizou, J.C.B. 2002 MNRAS 333, L26.
\bibitem[]{} Ogilvie \& Lubow 2003 ApJ 587, 398.
\bibitem[]{} Papaloizou, J.C.B. 2003 CeMDA 87, 53.
\bibitem[]{} Papaloizou, J.C.B. \& Terquem, C. 2001 MNRAS 325, 221.
\bibitem[]{} Papaloizou, J.C.B. \& Terquem, C. 2002 MNRAS 332, L39.
\bibitem[]{} Papalouizou, J.C.B., Nelson, R.P. \& Masset, F. 2001 A\&A 366, 263.
\bibitem[]{} Peale, S.J. 1986 in Satellites, ed. J.A. Burns \& M.S. Matthews (Tucson: Univ. Arizona Press)
\bibitem[]{} Pickett, B.K., Majia, A.C., Durisen, R.H., Cassen, P.M., Berry, D.K., Link, R.P. 2003 ApJ 590, 1060.
\bibitem[]{} Pollack, J.B., Hubickyj, O., Bodenheimer, P., Lissauer, J.J., Podolak, M. \& Greenzweig, Y. 1996 Icarus 124, 62. 
\bibitem[]{} Rasio, F.A. \& Ford, E.B. 1996 Science 274, 954.
\bibitem[]{} Rice, W.K.M., Armitage, P.J. 2003 ApJ 598, 55.
\bibitem[]{} Rivera, E.J. et al.\ 2005 ApJ 634, 625. 
\bibitem[]{} Snellgrove, M.D., Papaloizou, J.C.B., Nelson, R.P. 2001 A\&A 374, 1092.
\bibitem[]{} Stepinsky, T.F., Malhotra, R., \& Black, D.C. 2000 ApJ 545, 1004.
\bibitem[]{} Strom, R.G., Malhotra, R., Ito, T., Yoshida, F., Kring, D.A. 2005 Science 572, 1847.
\bibitem[]{} Thommes, E.W., Duncan, M.J., Levison, H.F. 1999 Nature 402, 635.
\bibitem[]{} Thommes, E.W., Duncan, M.J., Levison, H.F. 2002 AJ 123, 2862.
\bibitem[]{} Trilling, D.E., Benz, W., Guillot, T., Lunine, J.I., Hubbard, W.B., Burrows, A. 1998 ApJ 500, 428
\bibitem[]{} Tsiganis, K., Gomes, R., Morbidelli, A., Levison, H.F. 2005 Nature 435, 459.
\bibitem[]{} Veras, D. Armitage, P.J. 2004 Icarus 172, 349.
\bibitem[]{} Ward, W.R. 1997 Icarus 126, 261.
\bibitem[]{} Weidenschillingm S.J. \& Marzari, F. 1996 Nature 384, 619.

\end{thebibliography}
\end{document}